\documentclass[letter]{jpsj2} 

\title{
Possible Verification of Tilted Anisotropic Dirac Cone 
in $\alpha$-(BEDT-TTF)$_2$ I$_3$
Using Interlayer Magnetoresistance
}

\author{Takao \textsc{Morinari}\thanks{
morinari@yukawa.kyoto-u.ac.jp
},
Takahiro \textsc{Himura},
and Takami \textsc{Tohyama}
}

\inst{
Yukawa Institute for Theoretical Physics, Kyoto University, Kyoto 606-8502, Japan
}

\abst{
It is proposed that the presence of a tilted and anisotropic Dirac cone
can be verified using the interlayer 
magnetoresistance in the layered Dirac fermion system,
which is realized in quasi-two-dimensional organic compound
$\alpha$-(BEDT-TTF)$_2$ I$_3$.
Theoretical formula for the interlayer magnetoresistance
is derived using the analytic Landau level 
wave functions and assuming local tunneling of electrons.
It is shown that the resistivity 
as a function of the azimuthal angle of the magnetic field
takes the maximum in the direction
of the tilt if anisotropy of the Fermi velocity of the Dirac cone 
is small.
The procedure is described to determine the parameters of the 
tilt and anisotropy.
}

\kword{
$\alpha$-(BEDT-TTF)$_2$I$_3$,
Dirac cone, magnetoresistance}

\newcommand{\be}{\begin{equation}}
\newcommand{\ee}{\end{equation}}
\newcommand{\bea}{\begin{eqnarray}}
\newcommand{\eea}{\end{eqnarray}}

\begin{document}
\maketitle

There is a group of condensed matter systems
in which low-lying properties of the conduction electrons
are described by a relativistic Dirac equation 
with the velocity of light replaced by the Fermi velocity.
Remarkable physical phenomena of such Dirac fermions are 
clearly demonstrated in graphene, a monolayer of graphite.
\cite{Novoselov2005,Zhang2005}
In the observed integer quantum Hall effect
the plateaus in the Hall conductivity are characterized by not integers, $N$,
but $N+1/2$.
The origin of the shift of $1/2$ is the presence of the zero energy Landau level
for the Dirac fermions.

In contrast to graphene, which is a purely two-dimensional system,
the organic conductor $\alpha$-(BEDT-TTF)$_2$ I$_3$\cite{Bender1984}
is the first bulk material where the 
massless Dirac fermion-like spectrum is realized.
Experimentally it was 
reported that this quasi-two-dimensional compound was a 
narrow gap system under high pressure.
\cite{Kajita1992,Tajima2000,Tajima2006,Tajima2007}
Using the tight-binding model with 
the transfer integrals obtained by X-ray diffraction 
experiment \cite{Kondo2005},
Kobayashi {\it et al.} 
suggested that the system is zero-gap and
the energy dispersion is linear 
around the zero-gap point.
\cite{Kobayashi2004,Katayama2006}
They showed that the electronic band structure is
described by a tilted and anisotropic Dirac cone.
The first principle calculations supported this Dirac cone 
structure.\cite{Ishibashi2006,Kino2006}
Although observed negative interlayer magnetoresistance\cite{Tajima2008u}
is an evidence of the presence of a Dirac cone,
as theoretically explained by Osada\cite{Osada2008}
in terms of the zero energy Landau level wave function,
there is no direct experimental verification
that the cone is tilted and anisotropic.

In this Letter, we propose that the Dirac cone structure 
is verified by analyzing interlayer magnetoresistance
through its dependence on the applied magnetic field direction.
Using the analytic form of the zero-energy Landau level wave function
with tilt and anisotropy of the Dirac cone,\cite{Goerbig2008}
we derive a formula for the interlayer magnetoresistance.
Our formula is unique for the analysis of the tilted
and anisotropic Dirac cone in $\alpha$-(BEDT-TTF)$_2$I$_3$.
It should be noted that
in $\alpha$-(BEDT-TTF)$_2$I$_3$ 
Shubnikov-de Haas oscillations, which is a standard experiment
to study an electronic structure,
have never been observed and
angle resolved photoemission spectroscopy,
which directly confirmed the linear dispersion in graphene,\cite{Zhou2006}
is not applicable to organic compounds.

\begin{figure}
   \begin{center}
    \includegraphics[width=3.2in]{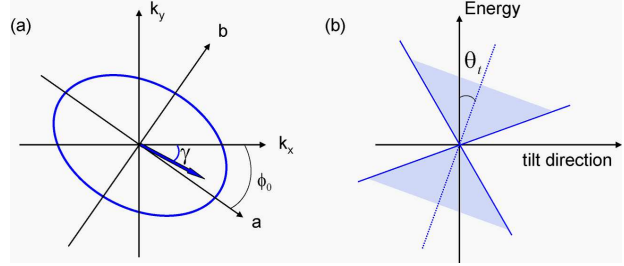}
   \end{center}
   \caption{ \label{fig_parameters}
	(a)The definition of the crystal axes denoted by $a$ and $b$ 
	and the wave vector axes $k_x$ and $k_y$.
	The angle between the $a$-axis and the $k_x$-axis is 
	represented by $\phi_0$.
	The arrow indicates the tilt direction.
	The angle between the $k_x$-axis and the tilt direction
	is defined as $\gamma$.
	By tilting the cone the intersection becomes an ellipse
	whose long axis is in the tilt direction as schematically
	shown in the figure.
	(b)The definition of the tilt angle, $\theta_t$.
	The horizontal axis is the tilt direction in the $k_x$-$k_y$ plane.
	The vertical axis is the energy axis.
        The declined solid lines represent the dispersion of the Dirac cone
        along the tilt direction.
	The dotted line represents the center line of the two dispersions.
	The angle between the energy axis and the center line is $\theta_t$.
    }
 \end{figure}

Before beginning the analysis of the system, we introduce
parameters defining the tilt and anisotropy 
of the Dirac cone.
We represent the crystal axes in the plane as $a$ and $b$.
We assume that the intersections of an anisotropic Dirac cone are elliptic.
The principal axes are not necessarily parallel to the crystal axes.
So the $k_x$- and $k_y$-axes in the wave vector space are taken so that
those axes are parallel to the principal axes.
The angle between the $k_x$-axis and the $a$-axis 
is denoted by $\phi_0$ as shown in Fig.\ref{fig_parameters}(a).
As for the tilt direction, we introduce $\gamma$ to denote 
the angle between the $k_x$-axis and the tilt direction.
The parameter $\theta_t$ describes the tilt angle
as defined in Fig.\ref{fig_parameters}(b).
Thus, the parameters are $\gamma$, $\theta_t$, and
the Fermi velocity anisotropy (denoted by $\alpha$ below).
Within our formulation,
$\phi_0$ is taken as a given parameter.

Now we start with the Hamiltonian describing a single layer 
of an anisotropic and tilted Dirac cone system
following Ref.\citenum{Kobayashi2007},
\be
H = \sum\limits_{k_x ,k_y } {H\left( {k_x ,k_y } \right)},
\ee
where
\be
H\left( {k_x ,k_y } \right) = \hbar \left( {\begin{array}{*{20}c}
   {v_0^x k_x  + v_0^y k_y } & {v_x k_x  - iv_y k_y }  \\
   {v_x k_x  + iv_y k_y } & {v_0^x k_x  + v_0^y k_y }  \\
\end{array}} \right).
\ee
This is a sufficiently general form.
Anisotropy in the Fermi velocity is parameterized by 
$\alpha = \sqrt{v_x/v_y}$.
In the absence of the magnetic field
the energy dispersion is 
given by $\epsilon_k = \hbar \left( v_0^x  k_x + v_0^y k_y \pm 
\sqrt{v_x^2 k_x^2 + v_y^2 k_y^2} \right) $.
The angle $\gamma$ defined in Fig.\ref{fig_parameters}(a) is given by 
\be
\gamma = \cot^{-1} \frac{v_0^x v_y}{v_0^y v_x }.
\ee

To compute the interlayer magnetoresistance, we need 
the Landau level wave functions.
We represent the magnetic field as
$(B_x, B_y, B_z)=B(\cos \theta \cos \phi, \cos \theta \sin \phi, \sin \theta)$.
Here $\phi$ is the angle in the plane with respect to the positive $k_x$-axis
and $\theta$ is the angle between the magnetic field direction and the plane.
We choose the gauge so that the vector potential 
is given by $A_x = B_y z + A_x^{(z)}$, $A_y = -B_x z + A_y^{(z)}$, and $A_z=0$
where $B_z = \partial_x A_y^{(z)} - \partial_y A_x^{(z)}$.
The presence of the inplane magnetic field is taken into account
by a gauge transformation because they depend only on $z$.
The zero-energy Landau level wave function 
is given in Ref.\citenum{Goerbig2008}.
Here we derive it in a different way which can be applicable
to compute other Landau level wave functions.
The derivation procedure consists of three steps.
After the Pierls substitution $(k_x, k_y) \rightarrow
(\kappa_x, \kappa_y) = (k_x + e A_x^{(z)}/c \hbar, k_y + e A_y^{(z)}/c \hbar)$,
first we rescale $k_y$ so that the Fermi velocity 
in this direction is $v_x$.
At this transformation, $B_z$ is multiplied by $v_y/v_x$.
Secondly we rotate the system by the angle $\gamma$ in the plane.
The transformed Hamiltonian is,
\be
H\left( {\kappa _x ,\kappa _y } \right) 
= \hbar v_x U^\dag  \left( { - \eta \kappa _x \sigma _0  
+ \kappa _x \sigma _1  + \kappa _y \sigma _2 } \right)U,
\ee
where $\kappa_x$ and $\kappa_y$ are redefined 
as the rotated variables, $\sigma_j$ ($j=1,2$) are the Pauli matrices 
and
\be
\eta  =  
\sqrt {\left( {\frac{{v_0^x }}{{v_x }}} \right)^2  
+ \left( {\frac{{v_0^y }}{{v_y }}} \right)^2 }, 
\ee
\be
U = \left( {\begin{array}{*{20}c}
   {\exp({\frac{i}{2}\gamma }) } & 0  \\
   0 & {\exp({ - \frac{i}{2}\gamma }) }  \\
\end{array}} \right).
\ee
Now the direction of tilt is in the $\kappa_x$-axis.
The angle of tilt is defined by 
$\theta_t = \frac{\theta_+ - \theta_-}{2}$,
where $\tan \theta_{\pm} = 1 \pm \eta$.
In the Schr{\" o}dinger equation with the 
Hamiltonian $U H U^{\dagger}$,
we subtract the term associated with the tilt
from the both side of the equation.
Applying the operator of the Dirac cone and after some algebra
we obtain the operator form of an anisotropic harmonic oscillator.
Diagonalizing the operator, we find the energies of the Landau levels,
\be
\epsilon _n
= {\rm sgn} (n)
\left( {\hbar v_x /\ell _z } \right)\sqrt {2\lambda ^3 |n|} 
\ee
with $n=0,\pm1,\pm2,...$,
the magnetic length $\ell_z=\sqrt{c\hbar/eB_z}$ and
\be
\lambda  = \sqrt{1-\eta^2} = 
\sqrt {1 - \left( {\frac{{v_0^x }}{{v_x }}} \right)^2  
- \left( {\frac{{v_0^y }}{{v_y }}} \right)^2 }.
\ee
The Landau level wave functions are obtained by taking the Landau gauge
for $(A_x^{(z)},A_y^{(z)})$,
\bea
\phi _n \left( {\zeta _k^{(n)} } \right) 
&=& \frac{1}{{\sqrt {4\left( {1 + \lambda } \right)} }}
\left[ \left( {1 - \delta _{n,0} } \right)\left( {\begin{array}{*{20}c}
   {1 + \lambda }  \\
   {\eta }  \\
\end{array}} \right) f_{\left| n \right| - 1} \left( {\zeta _k^{(n)} } \right) 
\right. \nonumber \\
& & \left. + {\mathop{\rm sgn}} \left( n \right)\left( {\begin{array}{*{20}c}
   { \eta }  \\
   {1 + \lambda }  \\
\end{array}} \right)f_{\left| n \right|} \left( {\zeta _k^{(n)} } \right)
\right],
\eea
with ${\rm sgn}(0)=1$.
Here $k$ is the wave vector for the plane wave component,
where the plane wave part is implicitly included, and
\be
\zeta _k^{(n)}  = \sqrt{\lambda}
\left[ 
\frac{\alpha}{\ell _z }\left( {x\sin \gamma  + y\cos \gamma } \right) 
+ \frac{\ell _z}{\alpha} k \right]
-\eta \sqrt{2|n|}{\rm sgn}(n),
\ee
\be
f_n \left( \zeta  \right) = \frac{{\left( { - 1} \right)^n }}{{2^{n/2} 
\pi ^{1/4} \sqrt {n!} }}H_n \left( \zeta  \right)
\exp \left( { - \frac{1}{2}\zeta ^2 } \right),
\ee
with $H_n(\zeta)$ the Hermite functions.

%

The interlayer conductivity is calculated using the Kubo 
formula.
As for the impurity scattering, 
we assume that the scattering leads to Lorentzian-shape
density of states with half value width of a constant $\Gamma$.
Although a self-consistent Born approximation shows that $\Gamma$ 
is magnetic field dependent,\cite{Shon1998} here we ignore field dependence
of $\Gamma$ for simplicity.
We focus on the contribution from the zero energy Landau level 
at zero temperature.
We shall comment on the effect of the other Landau levels later.
The matrix elements of the current operator is calculated
similarly to the non-tilted case \cite{Osada2008}
assuming local tunneling of electrons between two neighboring layers.
\cite{Osada2006}
The interlayer magnetoresistance is given by
\be
\rho _{zz}^{\left( 0 \right)}  
= \frac{A}{{B_0  + B\sin \theta 
\exp \left[ { - \frac{1}{2}\left( {\frac{{a_c }}{{\ell _z }}} \right)^2 
\frac{{\cos ^2 \theta }}{{\sin ^2 \theta }}
I\left( {\phi ,\alpha ,\gamma ,\lambda } \right)} \right]}},
\label{eq_rho_zz}
\ee
where 
\bea
I \left( {\phi ,\alpha ,\gamma ,\lambda } \right)
&=& \lambda \left( {\alpha \sin \phi \cos \gamma
- \frac{1}{\alpha }\cos \phi \sin \gamma } \right)^2  
\nonumber \\
& & \hspace{-4em}
+ \frac{1}{\lambda }\left( {\alpha \sin \phi \sin \gamma
+ \frac{1}{\alpha }\cos \phi \cos \gamma } \right)^2.
\eea
Here $B_0$ is a parameter of the theory and 
$A = \frac{\hbar }{{e^2 }}\left( 
{\frac{{2\pi ^2 \Gamma ^2 }}{{N_c t_c^2 }}} \right)
\left( {\frac{{c\hbar }}{{ea_c^2 }}} \right)$ is taken to be a constant
with $N_c$ the number of layers, $t_c$ the hopping parameter
between neighboring layers, and $a_c$ the lattice constant 
perpendicular to the plane.
Note that the function $I(\phi ,\alpha ,\gamma ,\lambda )$
has the period $\pi$ with respect to $\phi$ as
confirmed analytically.
Note also that there is another Dirac cone with 
$\eta \rightarrow -\eta$ by symmetry.\cite{Kobayashi2007}
But that Dirac cone has the same contribution because
the expression is invariant under this sign change of $\eta$.

\begin{figure}
   \begin{center}
    \includegraphics[width=3in]{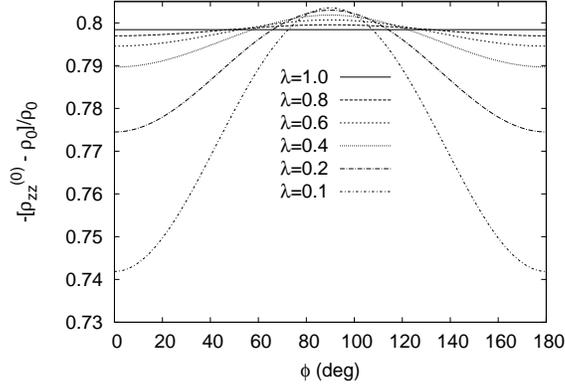}
   \end{center}
   \caption{ \label{fig_lambda}
	Dependence of the interlayer magnetoresistance on $\phi$
	for various tilt parameters $\lambda$
	at $B=6{\rm T}$, $B_0=0.5{\rm T}$ 
	and $\theta=20$ degrees with $\alpha=1$($v_x=v_y$).
    }
 \end{figure}

\begin{figure}
   \begin{center}
    \includegraphics[width=3in]{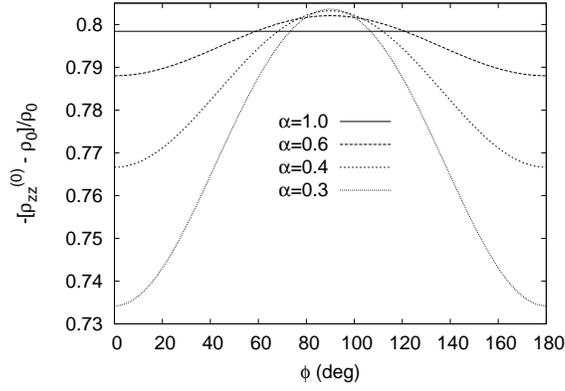}
   \end{center}
   \caption{ \label{fig_alpha}
	Dependence of the interlayer magnetoresistance on $\phi$
	in the absence of the tilt ($\lambda=1$)
        for various $\alpha$ values.
	Other parameters are the same as in Fig.\ref{fig_lambda}.
    }
 \end{figure}

Let us move on to the parameter dependence of the interlayer 
magnetoresistance.
Figure \ref{fig_lambda} shows the angular $\phi$ dependence 
of $\delta \rho_{zz}^{(0)} = - [\rho_{zz}^{(0)}-\rho_0]/\rho_0$
for various values of $\lambda$ with $\alpha=1$($v_x=v_y$).
Here $\rho_0$ is the interlayer resistance in the absence 
of the magnetic field.
On the other hand, 
Fig. \ref{fig_alpha} shows the angular $\phi$ dependence 
for various values of $\alpha$ in the absence of the tilt.
Anisotropy of the Fermi velocity also leads to
an angular dependence of the interlayer magnetoresistance.
As for the $\theta$ dependence,
we found that the ratio of the maximum to the minimum 
in Figs. \ref{fig_lambda} and \ref{fig_alpha} decreases
with increasing $\theta$.


The fact that anisotropy in eq.(\ref{eq_rho_zz}) in the $x-y$ plane arises
from the tilt is understood as follows.
In the presence of tilt, the intersection
of the Dirac cone is deformed.
In case of $\eta=0$ and $v_x=v_y$, the intersection is circle.
But if $\eta \neq 0$, the intersection becomes an ellipse
with the origin at $(k_x^{(n)},k_y^{(n)})
=(\eta (\epsilon_n/\hbar v_x)/(1-\eta^2),0)$
and the ratio of the principal axes being $\lambda$.
The form of the Landau level wave functions
is deformed according to this change of the intersection.
We find that the Landau level wave function
is more localized in the tilt direction
than the perpendicular direction to the tilt direction.
Since the matrix element of the interlayer current operator has a Fourier 
transformation like form with respect to the in-plane magnetic field, 
the current operator matrix element takes the minimum if the in-plane magnetic field 
is in the tilt direction. Therefore, the interlayer resistivity takes the maximum 
if the in-plane magnetic field is in the tilt direction.

\begin{figure}
   \begin{center}
    \includegraphics[width=3in]{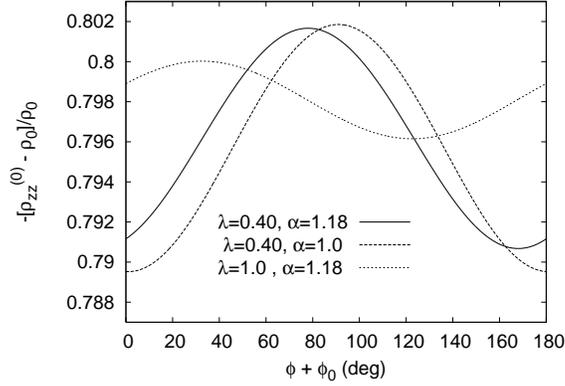}
   \end{center}
   \caption{ \label{fig_nagoya}
	Dependence of the interlayer magnetoresistance on $\phi$
	using the parameters estimated by Kobayashi {\it et al.}
	\cite{Kobayashi2007} at $P_a=4.5{\rm kbar}$.
	Other parameters are set as $B=6{\rm T}$, $B_0=0.5{\rm T}$,
	and $\theta=20$ degrees.
	The results for no tilting case and isotropic Fermi velocity case
        are shown as well.
    }
 \end{figure}

Kobayashi {\it et al.} estimated the parameters of the anisotropic 
tilted Dirac cone at the uniaxial pressure 
$P_a=4.5{\rm kbar}$ along the $a$-axis.\cite{Kobayashi2007}
Using those values, we find
$\gamma=-31.6$ degrees,
$\eta=0.92$, $\lambda=0.40$, $\alpha=1.18$,
and $\phi_0=32.6$ degrees.
Figure \ref{fig_nagoya} shows the interlayer magnetoresistance
for this parameter set.
From the comparisons with the no tilt case ($\lambda=1$, $\alpha=1.18$)
and the isotropic Fermi velocity case ($\lambda=0.40$, $\alpha=1.0$)
with using the same other parameter values,
we see that the $\phi$-dependent interlayer magnetoresistance
comes from the tilt of the Dirac cone.

Experimentally a rough estimation of the tilt angle, $\gamma$,
is obtained if we assume $\alpha \simeq 1$.
In this case anisotropy of the interlayer magnetoresistnace
in the plane is mainly associated with the tilt.
For $\alpha = 1$, the function $I$ takes a simple form,
\be
I\left( {\phi ,\alpha  = 1,\gamma ,\lambda } \right) 
= \frac{1}{2}\left( {\frac{1}{\lambda } + \lambda} \right) 
+ \frac{1}{2}\left( {\frac{1}{\lambda } - \lambda} \right)
\cos 2\left( {\phi  - \gamma } \right).
\label{eq_approx}
\ee
From this expression one can see that the resistivity takes 
the maximum in the direction of the tilt.
An approximate value of $\gamma$ is found from the angle of that direction.
The other parameters $\lambda$ and $\alpha$ can be estimated as follows.
We first extract the function 
$I \left( {\phi ,\alpha ,\gamma ,\lambda } \right)$ part
from the experimental data
using the parameters $A$ and $B_0$ 
determined from $\theta$ dependence of $\rho_{zz}^{\left( 0 \right)}$.
Taking the approximate values of $\gamma$ and $\lambda$,
which is obtained by using eq.(\ref{eq_approx}),
and $\alpha=1$ as initial values, more precise values are determined 
by the least squares method.

Now we comment on the conditions for the application of
the formula (\ref{eq_rho_zz}).
The formula is derived by using the zero-energy Landau level wave function.
To justify this approximation, 
the magnetic field $B_z$ should be large enough.
The energy gap to the first excited Landau level 
is $\epsilon_1 \simeq 40\sqrt{B_z}{\rm K}$ where $B_z$ is measured
in units of tesla if we assume the average Fermi velocity 
$\simeq 10^7 {\rm cm/s}$.\cite{Tajima2008u}
Therefore, if the temperature is sufficiently lower than
$\epsilon_1$ then the effect of the other Landau levels 
is safely neglected.
As for the angle $\theta$, the condition is 
$40\sqrt{B \sin \theta} > k_B T$.
The formula including the effect of the other Landau levels
is necessary for small $B_z$ values.
It is straightforward to extend the formula to
this case.
But the formula is complicated and such a formula
is necessary only when one tries to see the crossover 
from the quantum limit to the semi-classical regime.
However, this is beyond the scope of this paper.
Another effect to be concerned is the Zeeman splitting.
\cite{Osada2008}
However, the Zeeman splitting leads to a constant shift
of the energy even though the shift depends on the spin.
At fixed total magnetic field $B$
the shift is unimportant for the determination of the 
parameters of the Dirac cone
because it just leads to a modification of $A$ in eq.(\ref{eq_rho_zz}).

To conclude, we have derived the formula for the interlayer magnetoresistance
in the presence of the tilt and the Fermi velocity anisotropy 
of the Dirac cone.
The direction of the tilt is determined from the 
azimuthal angle dependence of the interlayer magnetoresistance.
If the interlayer resistivity takes the maximum 
in some direction, then the Dirac cone is tilted 
in that direction.
Physically the resistivity takes the maximum in the tilt direction because 
the interlayer current operator matrix element takes the minimum.
The derived formula can be used to extract pressure
dependence of the parameters of the tilted and 
anisotropic Dirac cone.
It would be interesting to see the difference between
the parameters determined by applying the formula to
analyze the experimental data and the band calculation
result.
We expect that discrepancy between them should be large 
under low pressures because the Dirac fermions become
unstable due to the electron correlation leading to
the charge ordering\cite{Kino1995,Seo2000,Takano2001,Wojciechowski2003} 
and/or superconductivity.\cite{Tajima2002,Kobayashi2004}

We would like to thank N. Tajima and C. Hotta for helpful discussions
and T. Osada for sending us Ref.\citenum{Osada2008} prior to publication.
We also thank M. Nakamura for useful comments.
This work was supported by the Grant-in-Aid for Scientific Research
from the Ministry of Education, Culture, Sports, Science and Technology (MEXT) 
of Japan, the Global COE Program 
"The Next Generation of Physics, Spun from Universality and Emergence," 
and Yukawa International Program for Quark-Hadron Sciences at YITP.


\bibliography{../../../references/tm_library2}


\end{document}